\documentclass[letterpaper, 10 pt, conference]{ieeeconf}

\usepackage{graphicx} 
\usepackage[T1]{fontenc}
\usepackage[latin9]{inputenc}
\usepackage{xcolor}
\usepackage{amsmath}
\usepackage{amssymb}
\usepackage{stackrel}
\usepackage[letterpaper, top=20.1mm, bottom=25.4mm, left=15.9mm, right=15.9mm]{geometry}
\IEEEoverridecommandlockouts 

\newtheorem{definition}{Definition}
\newtheorem{theorem}{Theorem}
\newtheorem{remark}{Remark}
\newtheorem{assumption}{Assumption}
\newtheorem{corollary}{Corollary}
\makeatletter

\begin{document}

\title{Open-loop control design for contraction in affine nonlinear systems}
\author{Mohamed Yassine Arkhis$^{1}$, Denis Efimov$^{1}$\thanks{\noindent$^{1}$Inria, Univ. Lille, CNRS, UMR 9189 - CRIStAL, F-59000
Lille, France.\protect }}

\maketitle

\begin{abstract}
    In this paper, first, it is shown that if a nonlinear time-varying system is contractive, then it is incrementally exponentially stable. Second, leveraging this result, under mild restrictions, an approach is proposed to design feedforward inputs for affine in control systems providing contraction/incremental exponential stability. Unlike standard stability notions, which have well-established control design techniques, this note can be considered among the first ones to provide such a tool for a kind of incremental stability. The theoretical findings are illustrated by examples.
\end{abstract}
\section{Introduction}
The notions of contraction and incremental stability are concepts that aim to study the behavior of the trajectories of a given system toward each other, without needing an attractor such as an equilibrium point. These concepts were found useful and natural in studying many complex biological and technical systems \cite{izhikevich2007dynamical}, \cite{pikovsky2007synchronization}.\\Contraction theory for the analysis of dynamical systems was first introduced in the work \cite{demidovich1961dissipativity}. Then, a region of contraction is a domain of the state space, on which the symmetric part of the Jacobian matrix of the system is uniformly negative definite with respect to some Riemannian metric \cite{lohmiller1998contraction}. It has been shown that the trajectories that start and stay in a contraction region remain close to each other, relative to the distance between initial conditions, and converge exponentially towards each other. \\Incremental stability notions, asymptotic and exponential, were formally introduced in the work \cite{forni2013differential}, where sufficient Finsler Lyapunov conditions, which are milder than the uniform negativity property of the Jacobian, have been provided. \\For a nonlinear system \begin{equation}\label{1}
    \dot{x}(t)=f(t,x(t)),\;t\geq t_0\in\mathbb{R},\;x(t)\in\mathbb{R}^n,
\end{equation}where \(f:\mathbb{R}\times \mathbb{R}^n\to \mathbb{R}^n\) is a class \(\mathcal{C}^1\) vector field, these conditions aim at proving that the origin of the displacement dynamics (the linearized dynamics of \eqref{1} along its trajectories):\begin{equation}\label{2}
    \dot{\delta x}(t)= J_f(t,x(t))\delta x(t)
\end{equation}where \(\delta x(t)\in\mathbb{R}^n\) is the displacement vector, \(J_f(t,x):=\frac{\partial f}{\partial x}(t,x)\) is the Jacobian of \(f\) in the state, and \(x(\cdot)\) is a solution of \eqref{1}, is uniformly globally asymptotically/exponentially stable. Uniformity is understood in the sense that the solution upper estimate is independent of the trajectory along which the system \eqref{1} is linearized. In \cite{forni2013differential} and \cite{lohmiller1998contraction}, it has been shown that the uniform negativity assumption on the Jacobian and the existence of an exponential Finsler Lyapunov function are both, under some smoothness condition on the vector field, sufficient conditions for the incremental exponential stability of the system \eqref{1}. \\The uniform negativity assumption on the Jacobian with respect to a Riemannian metric \(M(t,x)\) means that \begin{equation}\label{quadratic metric}(t,x,\delta x)\mapsto \delta x^{\top} M(t,x) \delta x\end{equation} is a Lyapunov function for the system \eqref{2}, where \(M(\cdot,\cdot)\) is a symmetric uniformly positive definite function \cite{lohmiller1998contraction}, that is \(\exists m>0\) such that \(M(t,x)\ge mI_n,\ \forall t\in\mathbb{R},\ x\in\mathbb{R}^n\). The Finsler Lyapunov condition (can also be called a contraction metric) generalizes the quadratic Lyapunov function \eqref{quadratic metric} to a generic one \cite{forni2013differential}.\\The analysis of the \textit{linear} time-varying system \eqref{2} can be, in many cases as we will see in this paper, more straightforward than directly analyzing the nonlinear system \eqref{1}, and, in some sense, more advantageous. Indeed, notice that the system \eqref{1} and the system:\begin{equation}
    \dot{x}=f(t,x)+ g(t)
\end{equation}where \(g:\mathbb{R}\to \mathbb{R}^n\) is any piecewise continuous function, have the same Jacobian.\\ It has been proven in \cite{andrieu2016transverse} that, in the case of autonomous systems \eqref{1} with a uniformly bounded Jacobian, and with the system \eqref{2} admitting a uniformly globally exponentially stable origin, the system \eqref{1} is, in fact, incrementally exponentially stable. In this paper, we prove this property for a generic time-varying system without any boundedness condition on the Jacobian.\\ Contrary to the various standard notions of Lyapunov stability, the notion of incremental stability still lacks in terms of control design techniques. To our knowledge, the only results addressing this issue are backstepping stabilizers such as \cite{zamani2013backstepping} where a state feedback is provided. It is important to highlight that in many cases the state is not available for the control design, and the open-loop entrainment frequently appears in nature for synchronization of different weakly coupled systems \cite{izhikevich2007dynamical}, \cite{pikovsky2007synchronization}. Thus, in the present study we will try to complement \cite{zamani2013backstepping} by considering the synthesis of feedforward strategies enforcing contraction. As it has been explained above, there are two main approaches to analyzing the incremental exponential stability of a system: either directly estimating the distance between two arbitrary solutions of \eqref{1}, or first proving that the system is contractive 
 by analyzing the uniform global exponential stability of the origin of \eqref{2} along an arbitrary solution of \eqref{1}, and subsequently concluding incremental exponential stability of the system. In this paper, using the second approach, we provide a sufficient condition for the existence of a feedforward input forcing a nonlinear system to be contractive/incrementally exponentially stable. We also provide a design synthesis of such an input and explain the intuition behind it. Finally, we discuss the advantages of the second approach compared to the first one, and its disadvantages, and showcase them through examples and simulation.\\
\textbf{Notation}
\begin{itemize}
    \item \(\mathbb{R}\) is the set of real numbers, and \(\mathbb{R}^*_+\) is the set of strictly positive real numbers.
    \item \(|\cdot|\) is the Euclidean norm on \(\mathbb{R}^n\).
    \item For a symmetric matrix \(P^{\top}=P\in\mathbb{R}^{n\times n},\) the minimal and maximal eigenvalues are denoted \(\lambda_{\min}(P)\) and \(\lambda_{\max}(P)\), respectively.
    \item A function \(u:\mathbb{R}\to \mathbb{R}\) is said to be of class \(\mathcal{C}^p\) on a compact set \(S\subset \mathbb{R}\) if its restriction on \(S\) can be extended to a class \(\mathcal{C}^p\) function on an open neighborhood of \(S\).
    \item For a set \(D\subset \mathbb{R}^n\), \(int(D)\) denotes its interior.
    \item for two symmetric matrices \(P,\ Q\in\mathbb{R}^{n\times n}\), we denote by \(P \ge Q\) the property:\begin{equation*}
        x^{\top}Px \ge x^{\top}Qx,\ \forall x\in\mathbb{R}^n.
    \end{equation*}
    \item \(I_n\) denotes the identity matrix of order \(n\).
\end{itemize}

\section{Preliminaries}
Consider a nonlinear system: \begin{equation}\label{8}
    \dot{x}=f(t,x)
\end{equation}where \(x\in\mathbb{R}^n\) is the state vector, \(t\in\mathbb{R}\), \(f:\mathbb{R}\times\mathbb{R}^n\to\mathbb{R}\) is a class \(\mathcal{C}^2\) vector field. And its linearized dynamics along a solution \(x(\cdot)\) has the form:\begin{equation}\label{9}
    \dot{\delta x}=J_f(t,x) \delta x.
\end{equation}where \(\delta x\in\mathbb{R}^n\) and \(J_f\) is the Jacobian of \(f\) in \(x\). Since we are studying incremental stability, we need to further assume that the system \eqref{8} is forward complete, i.e., \(\forall t_0\in \mathbb{R},\ \forall x_{t_0}\in\mathbb{R}^n\), a solution \(\phi_{t_0}(t,x_{t_0})\) of \eqref{8} exists for all \(t\ge t_0\).
\begin{remark}
    In the literature concerning incremental stability, see e.g. \cite{forni2013differential}, the vector field \(f\) is often taken of class \(\mathcal{C}^2\). Such a restriction is needed to guarantee that the solutions of \eqref{8} are of class \(\mathcal{C}^2\), when we can effectively use the linearized dynamics \eqref{9} to analyze the incremental stability of a system \eqref{8}, as we will do in the proof of Theorem \ref{thm 1}.
\end{remark}
\begin{definition}
    Given a number \(T>0\), we say that the vector field \(f\) is \(T\)-\(periodic\) if it holds that:\vspace{-5pt}\begin{equation}
        f(t+T,x)=f(t,x),\ \forall t\in\mathbb{R},\ x\in\mathbb{R}^n.
    \end{equation}
\end{definition}
\begin{definition}\label{def 2}
    The system \eqref{8} is contractive if there exist \(k,\lambda>0\) such that for any solution \(x(\cdot)\) of \eqref{8}, and any solution \(\delta x(\cdot)\) of the linearized dynamics \eqref{9}: \vspace{-3pt}\begin{equation}\label{11}
        |\delta x(t)|\le k|\delta x(t_0)| e^{-\lambda(t-t_0)},\ \forall t_0\in\mathbb{R},\forall t\ge t_0.
    \end{equation}
\end{definition}Notice that for this definition, a contraction region is not specified, since global exponential stability of the origin of \eqref{9} can be satisfied even when there is no quadratic metric of the form \eqref{quadratic metric} decreasing along solutions globally.
\begin{definition}\label{def 3}
    The system \eqref{8} is called incrementally exponentially stable (IES) if there exist \(k,\lambda>0\) such that for any two solutions \(x_1(\cdot),\ x_2(\cdot)\) of \eqref{8} and \(\forall t_0\in\mathbb{R},\ \forall t\ge t_0\):\vspace{-3pt}\begin{equation}
        |x_1(t)-x_2(t)|\le k|x_1(t_0)-x_2(t_0)| e^{-\lambda(t-t_0)}.
\end{equation}
\end{definition}
\section{Problem statement}
This paper studies the incremental exponential stability and stabilization of a nonlinear time-varying system. First, we show that a contractive system is in fact IES. Second, we aim at designing a feedforward input that makes a nonlinear system contractive.\\For these purposes we will consider a nonlinear affine in control system \begin{equation}\label{6}
    \dot{x}(t)= f(t,x(t))+u(t)G(t,x(t)),
\end{equation}where \(x(t)\in\mathbb{R}^n\) is the state, \(t\in\mathbb{R}\), \(f:\mathbb{R}\times\mathbb{R}^n\to\mathbb{R}^n\), \(G:\mathbb{R}\times\mathbb{R}^n\to\mathbb{R}^n\) and \(u:\mathbb{R}\to\mathbb{R}\) are of class \(\mathcal{C}^2\). Define the linearized dynamics along a solution \(x(\cdot)\) of \eqref{6}:\begin{equation}\label{7}
    \dot{\delta x}(t)=(J_f(t,x(t))+u(t)J_G(t,x(t)))\delta x(t),
\end{equation}where \(\delta x(t)\in\mathbb{R}^n\) is the displacement vector, \(J_f(\cdot,\cdot):=\frac{\partial f}{\partial x}(\cdot,\cdot)\) and \(J_G(\cdot,\cdot):=\frac{\partial G}{\partial x}(\cdot,\cdot)\) are the Jacobian matrices of \(f\) and \(G\) in the state \(x\), respectively.
\begin{remark}
    The reason for considering a nonlinear affine in control system instead of a generic nonlinear system of the form:\begin{equation}\label{4}
    \dot{x}(t)=F(t,x(t),u(t))
    \end{equation}where \(F:\mathbb{R}\times \mathbb{R}^n\times \mathbb{R}^p\to \mathbb{R}^n\) is of class \(\mathcal{C}^2\), is that we aim at controlling the linearized dynamics, therefore, we need an understanding of how the input \(u\) influences the Jacobian matrix of the vector field \(F\).
\end{remark}
Finally, we denote for \(t\in \mathbb{R},\ x\in \mathbb{R}^n\) the symmetric parts of these Jacobians as:\begin{equation}
    \begin{aligned}
        A(t,x)&= J_f(t,x)^{\top}+J_f(t,x),\\R(t,x)&= J_G(t,x)^{\top}+J_G(t,x).
    \end{aligned}
\end{equation}
\section{Main results}
\subsection{Connection between definitions \ref{def 2} and \ref{def 3}}
The first result of this paper concerns the relation between a contractive system and an IES system:
\begin{theorem}\label{thm 1}
    If the system \eqref{8} is contractive, then it is IES.
\end{theorem}
\begin{proof}
    Let \(t_0\in\mathbb{R}\) and \(x_0,x_1\in\mathbb{R}^n\). Denote by \(\Gamma(x_0,x_1)\) the set of \(\mathcal{C}^1\) curves \(\gamma:[0,1]\to \mathbb{R}^n\) such that \(\gamma(0)=x_0\) and \(\gamma(1)=x_1\). Since the square path length between \(x_0\) and \(x_1\) is minimized by the path \(\theta(s):=sx_0+(1-s)x_1\), then:\begin{equation}\label{12}
        \inf_{\gamma \in\Gamma(x_0,x_1)}\int_0^1 \left|\frac{\partial \gamma}{\partial s}(s)\right|^2 ds=\int_0^1 \left|\frac{\partial \theta}{\partial s}(s)\right|^2 ds=|x_0-x_1|^2.
    \end{equation}From \eqref{12} we have \(\forall t\ge t_0\):\begin{equation}
        |\phi_{t_0}(t,x_0)-\phi_{t_0}(t,x_1)|^2=\inf_{\gamma \in\mathcal{S}(t,t_0,x_0,x_1)}\int_0^1 \left|\frac{\partial \gamma}{\partial s}(s)\right|^2 ds.
    \end{equation} where \(\mathcal{S}(t,t_0,x_0,x_1)=\Gamma(\phi_{t_0}(t,x_0),\phi_{t_0}(t,x_1))\). Fixing a \(t\ge t_0\), the map \(s\in [0,1] \mapsto \phi_{t_0}(t,\theta(s))\) is a \(\mathcal{C}^1\) curve in \(\mathcal{S}(t,t_0,x_0,x_1)\), thus:\begin{equation}\label{14}
        |\phi_{t_0}(t,x_0)-\phi_{t_0}(t,x_1)|^2\le \int_0^1 \left|\frac{\partial}{\partial s}\phi_{t_0}(t,\theta (s))\right|^2 ds.
    \end{equation}Since \(f\) is of class \(\mathcal{C}^2\), then \(\phi_{t_0}(\cdot,\cdot)\) is also \(\mathcal{C}^2\) (\cite{boothby2003introduction}, Theorem 4.1), as such, for a \(\mathcal{C}^2\) curve \(\gamma\in \Gamma(x_0,x_1)\), the map \((t,s)\in[t_0,+\infty)\times [0,1] \mapsto \phi_{t_0}(t,\gamma(s))\) satisfies:\begin{equation}
        \begin{aligned}
            \frac{\partial}{\partial t}\frac{\partial}{\partial s}\phi_{t_0}(t,\gamma(s))&=\frac{\partial}{\partial s}\frac{\partial}{\partial t}\phi_{t_0}(t,\gamma(s))\\&=\frac{\partial}{\partial s}f(t,\phi_{t_0}(t,\gamma(s)))\\&= J_f(t,\phi_{t_0}(t,\gamma(s)))\frac{\partial}{\partial s} \phi_{t_0}(t,\gamma(s)),
        \end{aligned}
    \end{equation}therefore the map \(t\in[t_0,+\infty)\mapsto \frac{\partial}{\partial s}\phi_{t_0}(t,\gamma(s))\) is a solution of \eqref{9}, as such:\begin{equation}
        \left|\frac{\partial}{\partial s}\phi_{t_0}(t,\theta(s))\right|\le k\left|\frac{\partial}{\partial s}\phi_{t_0}(t_0,\theta(s))\right| e^{-\lambda(t-t_0)},\ \forall t\ge t_0,
    \end{equation}since \(\phi_{t_0}(t_0,\theta(s))=\theta(s)\ \forall s\in[0,1]\), then:\begin{equation}
        \left|\frac{\partial}{\partial s}\phi_{t_0}(t,\theta(s))\right|\le k\left|\frac{\partial \theta}{\partial s}(s)\right| e^{-\lambda(t-t_0)},\ \forall t\ge t_0,
    \end{equation}combining this with \eqref{14} we obtain \(\forall t\ge t_0\):\begin{equation}
    \begin{aligned}
        |\phi_{t_0}(t,x_0)-\phi_{t_0}(t,x_1)|&\le k \left(\int_0^1 \left|\frac{\partial \theta}{\partial s}(s)\right|^2 ds\right)^{\frac{1}{2}}  e^{-\lambda(t-t_0)}\\&=k |x_0-x_1| e^{-\lambda(t-t_0)}\end{aligned}
    \end{equation}thus, the system \eqref{8} is IES.
\end{proof}The techniques in this proof have been inspired by the proofs of the main results of the papers \cite{forni2013differential}, \cite{kawano2024incremental}.
\subsection{Control design for contraction/incremental exponential stability}\label{section 4}
The second result concerns an open-loop control design for the contraction/incremental exponential stability of the system \eqref{6}.\\
Since our goal is to make the origin of the system \eqref{7} uniformly globally exponentially stable, and the dynamics of the squared norm of the system \eqref{7} is: \begin{equation}
    \frac{\partial}{\partial t} |\delta x(t)|^2= \delta x(t)^{\top}(A(t,x(t))+u(t)R(t,x(t)))\delta x(t),
\end{equation}then we need to find a sufficiently smooth input \(u(t)\) such that the matrix \(A(t,x)+u(t)R(t,x)\) is uniformly negative definite (uniform in state and time). However, this cannot be achieved in many cases. For instance, when \(A(t,x)>0\) and \(R(t,x)\) is sign-indefinite for some values of arguments. In these cases, overshoots (i.e., an increase in the norm of \(\delta x\) and therefore increase in the distance between solutions of \eqref{6}) are inevitable independently in the choice of \(u\). Fortunately, exhibiting overshoots does not imply lack of exponential total decay as long as the overshoots are compensated in other regions where \(A(t,x)+u(t)R(t,x)<0\). To handle these scenarios we need the following:
\begin{assumption}\label{assumption 1} \(\exists m>0\) and a strictly increasing radially unbounded sequence of time instants \((a_i)_{i\in\mathbb{Z}}\) such that:\begin{equation}\label{19}
    [a_{2i},a_{2i+1}]\subset \{t\in\mathbb{R}:\ \forall x\in\mathbb{R}^n:\ R(t,x)\ge mI_n\},
\end{equation}or:\begin{equation}\label{20}
    [a_{2i},a_{2i+1}]\subset \{t\in\mathbb{R}:\ \forall x\in\mathbb{R}^n:\ R(t,x)\le -mI_n\},
\end{equation}and \(\exists\ M>0,\  \varepsilon:\mathbb{R}\to\mathbb{R}_+^*\) such that:\begin{equation}\label{22}
A(t,x)\le MI_n,\ \forall t\in D^1,\ x\in \mathbb{R}^n,
\end{equation}\begin{equation}\label{21}
    A(t,x)\le \varepsilon(t)I_n,\ \forall t\in D^2,\ x\in\mathbb{R}^n,
\end{equation} where \(D^1=\cup_{i\in\mathbb{Z}}[a_{2i+1},a_{2i+2}]\), and \(D^2=\cup_{i\in\mathbb{Z}}[a_{2i},a_{2i+1}]\). And \(\exists k,L>0\) such that \(\forall i \in \mathbb{Z}\):\begin{equation}\label{23}
        a_{2i+1}-a_{2i}\ge k,
\end{equation}\begin{equation}\label{24}
    a_{2i+2}-a_{2i+1}\le L.
\end{equation}
\end{assumption}
Inequalities \eqref{19}, \eqref{20} and \eqref{24} show that the region where \(R(t,x)\) is sign-indefinite does not contain infinitely growing intervals, as such, it is possible to guide solutions outside this region. Inequality \eqref{21} implies that with a right input, \(u(t)R(t,x)\) can compensate for \(A(t,x)\), while inequality \eqref{22} means that even in the region where overshoots are inevitable, they are still bounded and therefore can be compensated by providing a sufficiently monotonous convergence in the other region. Finally, inequality \eqref{23} is needed to understand the minimum amount of monotonous convergence obtained in each interval of the region where we know \(R(t,x)\) is sign-definite.
\begin{remark}
    Notice that this assumption does not take into account whether \(A(t,x)\) is uniformly negative (uniform in the state) for some instants of time. Intuitively, one can see that in such a case, a small \(u\) for those times can work. Moreover, we can require the fulfillment of all above conditions only for the time instants when \(A(t,x) > -\alpha I_n\) for some \(\alpha>0\) defining the decay rate, and this requirement is omitted for brevity. Nevertheless, the input designed in the proof of Theorem \ref{thm 2} will still force the system to be contractive/IES.
\end{remark}
\begin{theorem}\label{thm 2}
    Under Assumption \ref{assumption 1}, there exists a class \(\mathcal{C}^2\) function \(u:\mathbb{R}\to\mathbb{R}\) that forces the system \eqref{6} to be contractive/IES.
\end{theorem}
\begin{proof}
    Denote \(D_{2i}=[a_{2i},a_{2i+1}],\ D_{2i+1}=[a_{2i+1},a_{2i+2}]\) and \(k_{2i}=\max\{\varepsilon(t):\ t\in D_{2i}\}\) for \(i\in \mathbb{Z}\). Let \(\alpha>0,\ c>\alpha+\frac{e^{(M+\alpha)L}}{k}\).\\We start by defining \(u\) on \(D_{2i+1}\), \(i\in\mathbb{Z}\). We distinguish two cases.\\ \textbf{\(1^{st}\) case}: \(\exists t\in D_{2i+1},\ \exists x\in \mathbb{R}^n\) such that \( \lambda_{\min}(R(t,x))=0\) or \(\lambda_{\max}(R(t,x))=0\).\\ If \(R(t,x)\ge mI_n\) on \(D_{2i}\), then we denote:\begin{equation}
        T_1^{2i+1}:=\min\{t\in D_{2i+1}: \exists x\in \mathbb{R}^n:\ \lambda_{\min}(R(t,x))=0\},
    \end{equation}if \(R(t,x)\le -mI_n\) on \(D_{2i}\), then:\begin{equation}
        T_1^{2i+1}:=\min\{t\in D_{2i+1}: \exists x\in \mathbb{R}^n:\ \lambda_{\max}(R(t,x))=0\},
    \end{equation} if \(R(t,x)\ge mI_n\) on \(D_{2i+2}\), then:\begin{equation}
        T_2^{2i+1}:=\max\{t\in D_{2i+1}: \exists x\in \mathbb{R}^n:\ \lambda_{\min}(R(t,x))=0\},
    \end{equation} if \(R(t,x)\le -mI_n\) on \(D_{2i+2}\), then:\begin{equation}
        T_2^{2i+1}:=\max\{t\in D_{2i+1}: \exists x\in \mathbb{R}^n:\ \lambda_{\max}(R(t,x))=0\}.
    \end{equation}These bounds exist due to the continuity of \(R(\cdot,\cdot)\).\\We define \(u(\cdot)\) on each \(D_{2i+1}\) as follows:\begin{itemize}
        \item on \([a_{2i+1},T_1^{2i+1}]\), \(u\) is any class \(\mathcal{C}^2\) function satisfying \(u(T_1^{2i+1})=0,\ u'(T_1^{2i+1})=0,\ u''(T_1^{2i+1})=0\), and -in the case of \( R(t,x)\ge mI_n\) on \(D_{2i}\):\begin{equation}
            \left\{\begin{aligned}&u(a_{2i+1})=-c_{2i+1}m,\\ &-c_{2i+1} m \le u(t) \le 0,\ \forall t\in [a_{2i+1},T_1^{2i+1}],\end{aligned}\right.
        \end{equation}-in the case of \( R(t,x)\le -mI_n\) on \(D_{2i}\):\begin{equation}
            \left\{\begin{aligned}&u(a_{2i+1})=c_{2i+1}m,\\ &0 \le u(t) \le c_{2i+1} m,\ \forall t\in [a_{2i+1},T_1^{2i+1}],\end{aligned}\right.
        \end{equation}where \(c_{2i+1}> \max\left(\frac{c+k_{2i}}{m^2},\frac{c+k_{2i+2}}{m^2}\right)\) is a positive constant.\\A function \(u\) satisfying these properties clearly exists, since \(R(t,x)\) is sign-definite \(\forall t\in(a_{2i+1},T_1^{2i+1})\) and all \(x\in\mathbb{R}^n\).
        \item on \([T_1^{2i+1},T_2^{2i+1}]\), \(u(t)=0\).
        \item on \([T_2^{2i+1},a_{2i+2}]\), \(u\) is any class \(\mathcal{C}^2\) function satisfying \(u(T_2^{2i+1})=0,\ u'(T_2^{2i+1})=0,\ u''(T_2^{2i+1})=0\), and -in the case of \( R(t,x)\ge mI_n\) on \(D_{2i+2}\):\begin{equation}
            \left\{\begin{aligned}&u(a_{2i+2})=-c_{2i+1}m,\\ &-c_{2i+1} m \le u(t) \le 0,\ \forall t\in [T_2^{2i+1},a_{2i+2}],\end{aligned}\right.
        \end{equation}-in the case of \( R(t,x)\le -mI_n\) on \(D_{2i+2}\):\begin{equation}
            \left\{\begin{aligned}&u(a_{2i+2})=c_{2i+1}m,\\ &0 \le u(t) \le c_{2i+1} m,\ \forall t\in [T_2^{2i+1},a_{2i+2}].\end{aligned}\right.
        \end{equation}
    \end{itemize}\textbf{\(2^{nd}\) case}: \(\forall t\in D_{2i+1},\ \forall x\in \mathbb{R}^n\), \(R(t,x)\) is sign-definite. Since \(R(\cdot,\cdot)\) is continuous, either \(\lambda_{\min}(R(t,x))>0,\ \forall t\in D_{2i+1},\ x\in\mathbb{R}^n\), or \(\lambda_{\max}(R(t,x))<0,\ \forall t\in D_{2i+1},\ x\in\mathbb{R}^n\) is satisfied. We only treat the first sub-case, the second follows the same reasoning.\\ The first sub-case can be true only if \(R(t,x) \ge mI_n,\ \forall t\in D_{2i}\cup D_{2i+2},\ \forall x\in \mathbb{R}^n\), therefore, we define \(u\) on \(D_{2i+1}\) as any \(\mathcal{C}^2\) function satisfying: \(u(a_{2i+1})=u(a_{2i+2})= -c_{2i+1}m\) and \(u(t)\le 0,\ \forall t\in D_{2i+1}\).\\Now let us define the control \(u\) on the intervals \(D_{2i}\). Notice that \(\forall i \in \mathbb{Z}\): \begin{equation}
        \begin{aligned}
            \min(|u(a_{2i+1})|,|u(a_{2i+2})|) >\max\left(\frac{c+k_{2i}}{m},\frac{c+k_{2i+2}}{m}\right),
        \end{aligned}
    \end{equation}therefore, to complete \(u\) on \(D_{2i}\), choose any \(\mathcal{C}^2\) function satisfying: \begin{equation}\label{34}
        \left\{\begin{aligned}&u(t)\le -\frac{c+k_{2i}}{m},\forall t\in 
        D_{2i},\ \text{if } R(\cdot,\cdot)\ge mI_n\text{ on }D_{2i},\\&u(t)\ge \frac{c+k_{2i}}{m},\forall t\in D_{2i},\ \text{if } R(\cdot,\cdot)\le -mI_n\text{ on }D_{2i},\end{aligned}\right.
    \end{equation}an example of such a function would be a polynomial in parts.\\The obtained \(u\) is \(\mathcal{C}^2\) on \(\mathbb{R}\), and satisfies: \begin{equation}
        \left\{\begin{aligned}&u(t)R(t,x) \le 0,\ \forall t\in D^1,\ x\in \mathbb{R}^n\\&u(t)\le -\frac{c+k_{2i}}{m},\forall t\in D_{2i},\ \text{if } R(\cdot,\cdot)\ge mI_n\text{ on }D_{2i},\\&u(t)\ge \frac{c+k_{2i}}{m},\forall t\in D_{2i}, \text{ if } R(\cdot,\cdot)\le -mI_n\text{ on }D_{2i}. \end{aligned}\right.
    \end{equation}
    Let us prove that this \(u(\cdot)\) makes the system IES. Consider a Finsler Lyapunov candidate: \(V(t,\delta x)=g(t)\delta x^{\top} \delta x\), where: \begin{equation}
            g(t)=\left\{\begin{aligned}
                &e^{-(M+\alpha)(t-a_{2i+1})}&&,\forall t\in D_{2i+1}\\&\xi_{2i+2}(t-a_{2i+2})+\zeta_{2i+2}&&, \forall t\in D_{2i+2}.
            \end{aligned}\right.
        \end{equation}for each \(i\in \mathbb{Z}\), where \(\xi_{2i+2}:=\frac{1-\zeta_{2i+2}}{a_{2i+3}-a_{2i+2}}\le \frac{1}{k}\) and \(\zeta_{2i+2}:=e^{-(M+\alpha)(a_{2i+2}-a_{2i+1})}\).\\We have: \begin{equation}
           e^{-(M+\alpha)L} \le g(t)\le 1,\ \forall t\in\mathbb{R},
        \end{equation}therefore, \(\forall t\in \mathbb{R},\ \forall \delta x\in\mathbb{R}^n\):\begin{equation}\label{38}
            e^{-(M+\alpha)L}|\delta x|^2 \le V(t,\delta x) \le |\delta x|^2 
        \end{equation}and, denoting \(g'(t):=\limsup_{s\to t^+}\frac{g(s)-g(t)}{s-t}\), we have:\begin{equation}
            \left\{\begin{aligned}
                &g'(t)\le (-M-\alpha) g(t)&&,\ \forall t\in int(D^1),\\&g'(t)\le \frac{1}{k}&&,\ \forall t\in D^2.
            \end{aligned}\right.
        \end{equation}The derivative of \(V\) along a trajectory \(\delta x(\cdot)\) of system \eqref{7} is:\begin{equation}
                \begin{aligned}
                \dot{V}(t,\delta x(t)):&=\limsup_{s\to t^+}\frac{V(s,\delta x(s))-V(t,\delta x(t))}{s-t}\\&\begin{aligned}=g'(t) \delta x(t)^{\top} \delta x(t)+ g(t) \delta x(t)^{\top} (A(t,x)\\+u(t)R(t)) \delta x(t).\end{aligned}
                \end{aligned}
        \end{equation}We have \(\forall t\in D_{2i}\):\begin{itemize}
        \item If \(R(t,x)\ge mI_n\) on \(D_{2i}\):\begin{equation}
        u(t)R(t)\le m u(t)I_n \le (-c-k_{2i})I_n,
        \end{equation}
        \item If \(R(t,x)\le -mI_n\) on \(D_{2i}\):\begin{equation}
            u(t)R(t)\le -m u(t)I_n \le (-c-k_{2i})I_n
        \end{equation}
        \end{itemize}and: \begin{equation}
            A(t,x)\le k_{2i},\ \forall t\in D_{2i}.
        \end{equation}Therefore, \(\forall t\in int(D^1)\):\begin{equation}
                \dot{V}(t,\delta x(t)) \le (-M-\alpha)V(t,\delta x(t))+M V(t,\delta x(t)),
        \end{equation}and \(\forall t\in D^2\):\begin{equation}
            \dot{V}(t,\delta x(t)) \le \frac{e^{(M+\alpha)L}}{k}V(t,\delta x(t))- c V(t,\delta x(t)).
        \end{equation}We conclude that for any solution \(\delta x(\cdot)\) of \eqref{7}, and \(\forall t\in \mathbb{R}\): \begin{equation}\label{46}
            \dot{V}(t,\delta x(t))\le -\alpha V(t,\delta x(t)).
        \end{equation}Combining \eqref{38} and \eqref{46}, we conclude from Theorem 4.10 of \cite{khalil2002control} that the system \eqref{6} is contractive and therefore IES by Theorem \ref{thm 1}.
\end{proof}
\begin{remark}
    Notice that the rate \(\alpha\) can be chosen arbitrarily. That is, a system satisfying Assumption \ref{assumption 1} can be forced to be contractive with any desired rate of exponential convergence.
\end{remark}\vspace{5pt}When the vector fields in \eqref{6} are periodic, Assumption \ref{assumption 1} can be significantly relaxed:
\begin{corollary}\label{cor 1}
    If \(f\) and \(G\) of system \eqref{6} are \(T\)-\(periodic\), and there exist \(m>0\) and a segment \([a,b]\subset [0,T]\) such that \(R(t,x)\ge mI_n,\ \forall t\in[a,b],\ x\in\mathbb{R}^n\) or \(R(t,x)\le -mI_n,\ \forall t\in[a,b],\ x\in\mathbb{R}^n\), then there exists a \(T\)-\(periodic\) input \(u\) that forces the system \eqref{6} to be contractive/IES, and approaching a \(T\)-\(periodic\) solution, i.e., \(\exists k,\lambda>0\) such that \(\forall t_0\in\mathbb{R}\), there exists a unique \(T\)-\(periodic\) solution \(\eta:[t_0,\infty)\to \mathbb{R}^n\) of \eqref{6} such that for any solution \(x\) of \eqref{6} defined on \([t_0,\infty)\), \(\forall t\ge t_0\):\vspace{-5pt}\begin{equation}
        |x(t)-\eta(t)|\le ke^{-\lambda(t-t_0)}|x(t_0)-\eta(t_0)|.
    \end{equation}
\end{corollary}
\begin{proof}
    Since \(f\) and \(G\) are \(T\)-\(periodic\), then \(A\) and \(R\) of the linearized dynamics \eqref{7} are also \(T\)-\(periodic\), and therefore Assumption \ref{assumption 1} is clearly satisfied. Furthermore, using the same design method of the proof of Theorem \ref{thm 2}, we determine a class \(\mathcal{C}^2\) input \(u\) on \([0,T]\),  such that the first-order derivatives of \(u\) at \(0^+\) and \(T^-\) are equal, as well as the second-order derivatives, so that we extend it to the whole \(\mathbb{R}\) to be a \(T\)-\(periodic\ \)\(\mathcal{C}^2\) input by definition, that forces the system to be contractive/IES. We conclude from Theorem 2 of \cite{russo2010global}, that the system is contractive/IES to a periodic solution.
\end{proof}
\section{Examples}
\subsection{Illustration for Corollary \ref{cor 1}}
Consider a simple scalar nonlinear control system of the form \eqref{6}:\begin{equation}\label{47}
    \dot{x}=x+u(t)\sin(t)x+g(t)
\end{equation}where \(g:\mathbb{R}\to\mathbb{R}\) can be any \(\mathcal{C}^2\) function. Then \(A(t,x)=2\) and \(R(t,x)=2\sin(t)\) clearly satisfy Assumption 1. In such a case, take as an example of \(D_{2i}\) the following: \(|\sin(t)|\ge \frac{1}{2},\ \forall t\in [\frac{\pi}{6}+2i\pi,\frac{5\pi}{6}+2i\pi]\cup [\frac{7\pi}{6}+2i\pi,\frac{11\pi}{6}+2i\pi], \forall i\in \mathbb{Z}\). Therefore, using the reasoning of the proof of Theorem \ref{thm 2}, we choose the input \(u\) satisfying inequality \eqref{34} on \(\cup_{i\in\mathbb{Z}}[\frac{\pi}{6}+2k\pi,\frac{5\pi}{6}+2k\pi]\cup [\frac{7\pi}{6}+2k\pi,\frac{11\pi}{6}+2k\pi]\) and with a different sign to \(\sin(\cdot)\) on its complement. Notice that in the design of a suitable \(u\), we ignore \(g\), thus, once \(u\) is found, we can choose \(g\) in any way we want, which in turn forces a sort of entrainement \cite{russo2010global}: trajectories will follow a certain shape that depends on \(u\) and \(g\).\\Perfoming simulations, Fig. \ref{1} illustrates the behavior of this system for \(u(t)=-3\sin(t)\) and \(g(t)=0\). Although there are overshoots, they are compensated enough in the other regions to attain contraction/incremental exponential stability.\\
\begin{figure}
\includegraphics[width=0.53\textwidth]{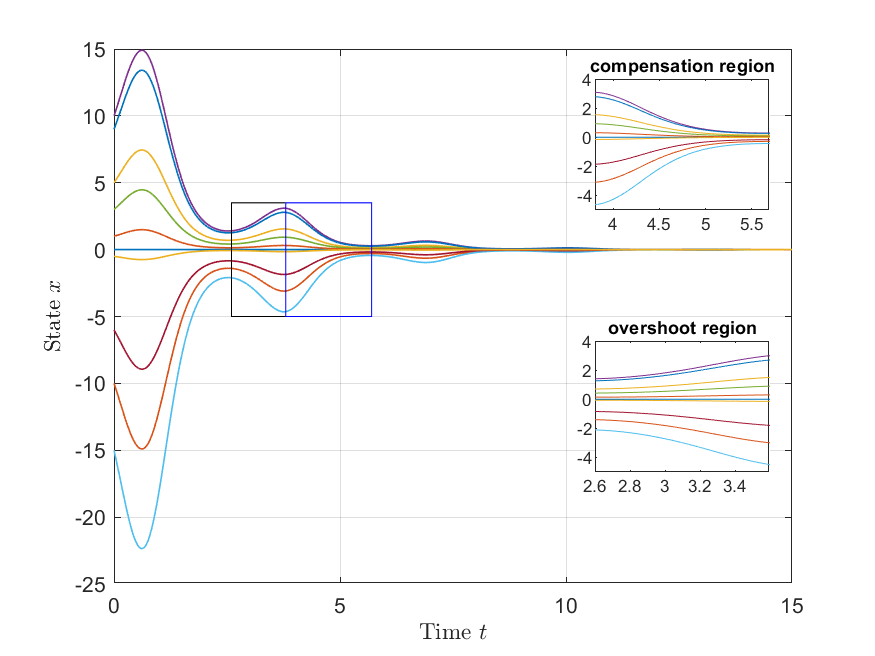}
\caption{Trajectories of the system \eqref{47} with \(u(t):=-3\sin(t)\) and \( g(t)\equiv 0\).}
\label{fig1}
\end{figure}Once a suitable \(u\) is designed, one can choose any \(g\), and the incremental exponential stability continues to hold, but the trajectory to which all solutions converge obviously changes, as shown in Fig. \ref{fig2}.
\begin{figure}
\includegraphics[width=0.53\textwidth]{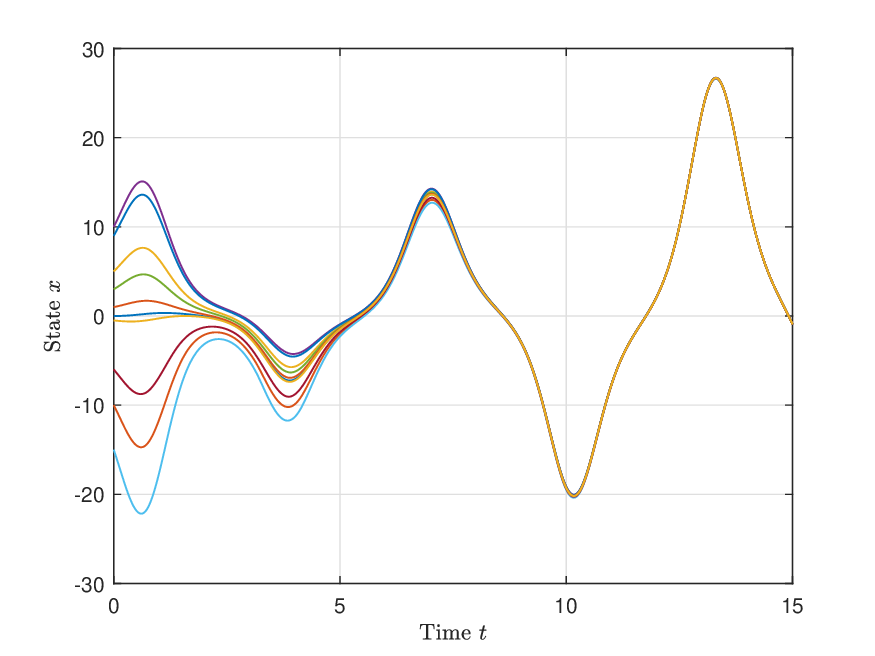}
\caption{Trajectories of the system \eqref{47} with \(u(t):=-3\sin(t)\) and \(g(t)=t \cos (t)\).}
\label{fig2}
\end{figure}
\subsection{Conservativeness of conditions}
\begin{figure}
\includegraphics[width=0.53\textwidth]{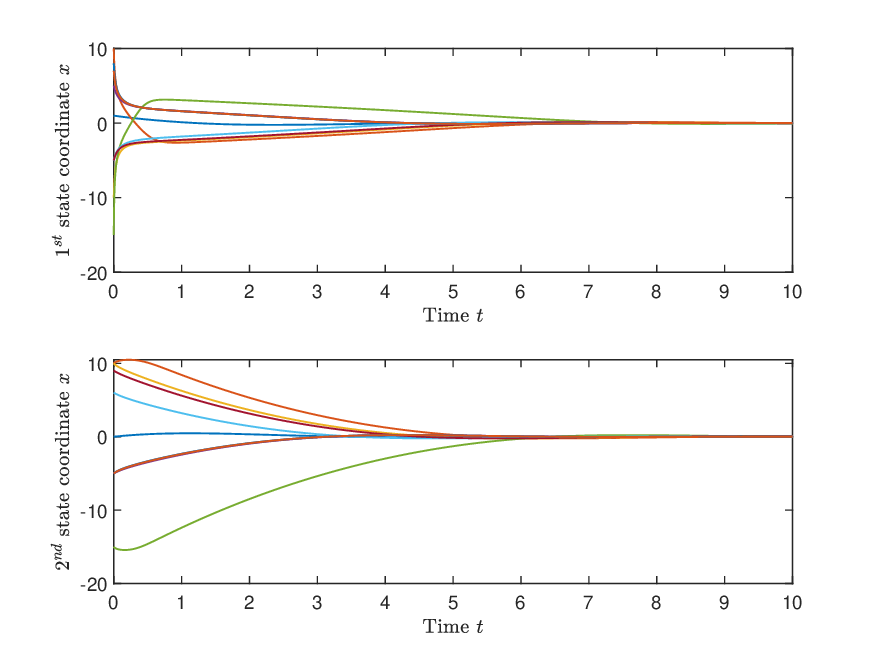}
\caption{Trajectories of the system \eqref{48} with \(u(t):=-2\).}
\label{fig3}
\end{figure}

\begin{figure}
\includegraphics[width=0.53\textwidth]{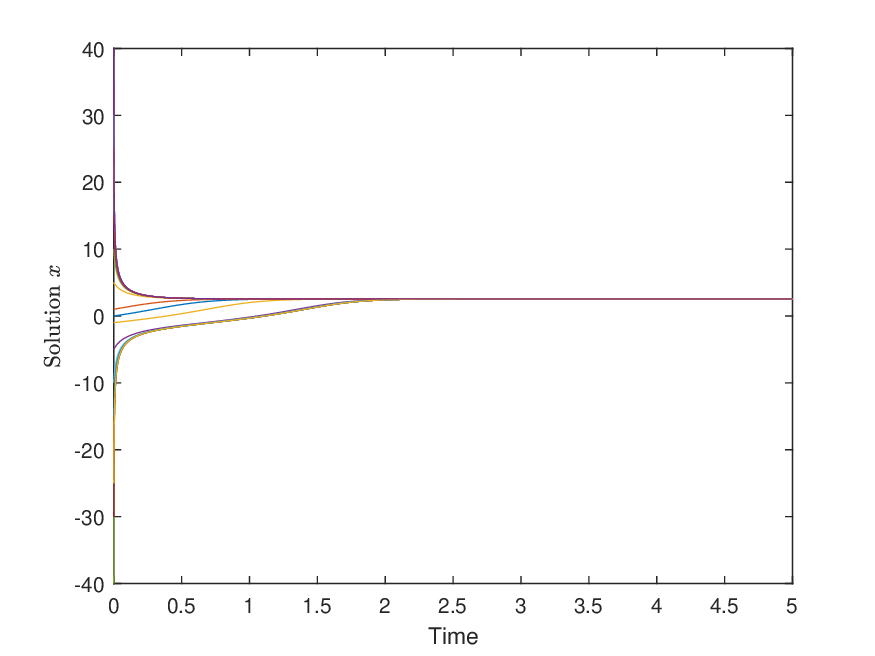}
\caption{Trajectories of the system \eqref{49} with \(u(t):=3\).}
\label{fig4}
\end{figure}

Although Assumption \ref{assumption 1} is very natural, it is clearly not generic. Among the excluded cases, some still allow the reasoning of the proof of Theorem \ref{thm 2} to remain applicable, while others render that reasoning ineffective.\\
For the first case, consider another simple nonlinear control system of the form \eqref{6}:\begin{equation}\label{48}
\left\{\begin{aligned}
    \dot{z}&=z-\frac{z^3}{3}-y+u(t)z\\ \dot{y}&=-y+z,\end{aligned}\right.
\end{equation}with \(x=\begin{bmatrix}
    z\\y
\end{bmatrix}\), then \(A(t,x)=2\begin{bmatrix} 
1-z^2 & 0 \\ 
0 & -1 
\end{bmatrix}\), and \(R(t,x)=\begin{bmatrix} 
2 & 0 \\ 
0 & 0 
\end{bmatrix}\). This example showcases a scenario that is not included in Assumption \ref{assumption 1}; the case where \(R(t,x)\) is always positive (or negative) semi-definite. If the eigenvalues of \(A(t,x)\) corresponding to the \(0\) eigenvalues of \(R(t,x)\) are always negative definite, then we can still apply the reasoning of the proof of Theorem \ref{thm 2}. Noticing that \(1-z^2<0,\ \forall |z|>1\), and \(0\le 1-z^2\le 1,\ \forall |z|\le 1\), in order to force \(A(t,x)+u(t)R(t,x)\) to be negative definite, we use a simple constant input \(u(t)= -2\) as shown in Fig. \ref{fig3}.\\For the case where that reasoning is no longer applicable, consider a scalar nonlinear control system: \begin{equation}\label{49}
    \dot{x}=x-\frac{x^3}{3} +u(t),
\end{equation}whose linearized dynamics along a solution \(x(\cdot)\) is: \begin{equation}
        \dot{\delta x}(t)=(1-x(t)^2)\delta x(t),
\end{equation}since \(u\) does not appear in the Jacobian, the reasoning of the proof of Theorem \ref{thm 2} is no longer valid. However, this does not mean that the system can not be forced to be IES, as can be seen in the Fig. \ref{fig4}.

\section{Conclusion}
This paper presents two main contributions. First, it establishes that for nonlinear time-varying systems, contractiveness implies incremental exponential stability. This naturally raises the question of whether the converse holds.\\Second, it introduces an open-loop control design method for achieving contraction and incremental exponential stability under mild assumptions. While these assumptions appear natural, they are not generic, leaving several open directions for future research. Specifically, if the input does not explicitly appear in the Jacobian, what conditions would enable the design of a suitable control? Furthermore, Is there a practical approach to addressing the incremental exponential stability problem directly through the main nonlinear system rather than through its linearized dynamics?\\ Regardless, one key takeaway remains: designing a feedforward control requires a certain level of understanding of the system's dynamics. The assumption in this paper focused on the notion of uniform global exponential stability, which allowed us to consider boundedness and convergence that is not necessarily monotonous. This leads to another important question: would it be more effective to base the design on an assumption that directly captures incremental stability? Whether such a condition could be sufficient for designing an appropriate feedforward control remains an open problem.
\bibliographystyle{plain}
\bibliography{bib}
\end{document}